\documentclass{pasj01}
\usepackage{color}
\Received{$\langle$reception date$\rangle$}
\Accepted{$\langle$acception date$\rangle$}
\Published{$\langle$publication date$\rangle$}
\begin{document}

\title{High-energy gamma-ray and neutrino production in star-forming galaxies across cosmic time: Difficulties in explaining the IceCube data}
\author{Takahiro \textsc{Sudoh},\altaffilmark{1,}$^{*}$
	Tomonori \textsc{Totani}\altaffilmark{1,2} and 
	Norita \textsc{Kawanaka}\altaffilmark{3,4}}
\altaffiltext{1}{Department of Astronomy, the University of Tokyo, Hongo, Tokyo 113-0033, Japan}
\altaffiltext{2}{Research Center for the Early Universe, 
the University of Tokyo, Hongo, Tokyo 113-0033, Japan}
\altaffiltext{3}{Department of Astronomy, Graduate School of Science, Kyoto University, Kitashirakawa Oiwake-cho, Sakyo-ku Kyoto, 606-8502, Japan}
\altaffiltext{4}{Hakubi Center, Kyoto University, Yoshida Honmachi, Sakyo-ku, Kyoto 606-8501, Japan}
\email{sudoh@astron.s.u-tokyo.ac.jp}

\maketitle

\begin{abstract}
We present a new theoretical modeling to predict luminosity and
spectrum of gamma-ray and neutrino emission of a star-forming galaxy,
from star formation rate ($\psi$), gas
mass ($M_{\rm gas}$), stellar mass, and disk size, taking into account
production, propagation and interactions of cosmic rays.  The model
reproduces the observed gamma-ray luminosities of nearby galaxies
detected by {\it Fermi} better than the simple power-law models as a
function of $\psi$ or $\psi M_{\rm gas}$. Then this model is used to
predict the cosmic background flux of gamma-ray and neutrinos from
star-forming galaxies, by using a semi-analytical model of
cosmological galaxy formation that reproduces many observed quantities
of local and high-redshift galaxies. Calibration of the model using
gamma-ray luminosities of nearby galaxies allows us to make a more
reliable prediction than previous studies. In our baseline model
star-forming galaxies produce about 20\% of isotropic gamma-ray
background unresolved by {\it Fermi}, and only 0.5\% of IceCube
neutrinos. Even with an extreme model assuming a hard injection
cosmic-ray spectral index of 2.0 for all galaxies, at most 22\% of
IceCube neutrinos can be accounted for.  These results indicate that
it is difficult to explain most of IceCube neutrinos by star-forming
galaxies, without violating the gamma-ray constraints from nearby
galaxies.
\end{abstract}

\section{Introduction}

The IceCube Collaboration has revealed the existence of
extraterrestrial high energy neutrinos, though their origin still
remains a mystery (\cite{IceCube13}). The arrival distribution is
consistent with being isotropic and the flavor ratio is consistent
with $\nu_e:\nu_{\mu}:\nu_{\tau} = 1:1:1$, suggesting an
extragalactic, astrophysical origin. A variety of source models has
been proposed so far, such as gamma-ray bursts, active galactic
nuclei, star-forming galaxies, and galaxy clusters, though no
individual source has been identified yet (for recent reviews, see,
e.g., \cite{Meszaros17}, \cite{Halzen17}).

Star-forming galaxies are one possible origin of the IceCube
neutrinos, in which cosmic rays (CRs) are produced by supernovae and
they produce pion-decay neutrinos via inelastic collisions with the
interstellar medium (ISM)
(\cite{Loeb06,Thompson06,Stecker07,Lacki11,Murase13,He13,Tamborra14,Anchordoqui14,Liu14,Emig15,Chang15,Giacinti15,Senno15,Chakraborty16,Xiao16,Bechtol17}).
Starburst galaxies are especially important in this context, because
CRs are expected to efficiently produce pions due to their high gas
densities and confinement by strong magnetic fields. However, a
definite conclusion has not yet been reached about whether star
forming galaxies are the dominant source of IceCube neutrinos.
\citet{Tamborra14} concluded that starburst sources have a possibility
to explain a significant fraction of the IceCube flux based on an
empirical relation between gamma-ray and infrared luminosities, while
\citet{Bechtol17} argued that galaxies cannot produce more than 30\%
of the IceCube data if the upper limit on non-blazar fraction of the
extragalactic gamma-ray background is considered.  \citet{Xiao16}
found that star-forming galaxies might explain about 50\% of the
IceCube flux, but assuming a dominant contribution from hypernovae,
which is rather uncertain.  In these studies, however, little
attention was paid to the consistency of the model with observed
gamma-ray fluxes from nearby star-forming galaxies
(\cite{Ackermann12SFG,Tang14,Peng16}), which should provide a useful
constraint because gamma-rays are inevitably emitted if neutrinos are
produced in a star-forming galaxy.

In this work, we present a new calculation of the contribution from
star-forming galaxies to the diffuse gamma-ray and
neutrino background, which takes into account cosmic-ray physical
processes in galaxies and is consistent with the gamma-ray
observations of nearby galaxies.  For this purpose, we construct a new
theoretical framework to predict both flux and spectrum of gamma-ray
and neutrino emissions from a galaxy based on star
formation rate (SFR), stellar mass, gas mass, and 
effective radius.

Gamma-ray flux from a galaxy has been modeled in a number of studies
to predict the cosmic gamma-ray background from star forming galaxies
(\cite{Strong76,Lichti78,Dar95,Pavlidou02,Thompson07,Ando09,Fields10,Makiya11,Stecker11,Ackermann12SFG,Chakraborty13,Lacki14,Komis17,Lamastra17}),
and most studies estimated gamma-ray luminosity only from one or two
physical quantities (e.g. SFR, gas mass, or infrared luminosity) assuming
empirical relations. However, such an approach may induce some bias in
predictions of the cosmic background \citep{Komis17}.  Gamma-ray
spectrum was also often treated in a simple way, using the Milky Way
and M82 spectra as templates for normal and starburst galaxies. Our
model includes a larger number of physical quantities of a galaxy to
take into account the production, propagation and interactions of
cosmic rays, and hence flux and spectrum can be calculated for diverse
individual galaxies across cosmic time.

This modeling is compared with the observed gamma-ray luminosities of
six nearby galaxies, and we will show that our modeling reproduces the
observed gamma-ray luminosities better than the simple scaling
relations with SFR and gas mass.  Then the cosmic background flux and
spectrum of gamma-rays and neutrinos are calculated by using a
semi-analytical model of cosmological galaxy formation (the Mitaka
model, \cite{Mitaka}), which reproduces many observational data of
local and high-$z$ galaxies. The Mitaka model provides us with the physical
quantities of galaxies to predict gamma-ray and neutrino emissions,
based on a self-consistent calculation of formation and evolution of
galaxies in a cosmological framework, taking into account key
baryonic processes such as gas cooling, star-formation, galaxy merger
and subsequent starbursts.

We organize this paper as follows.  The new model of gamma-ray and
neutrino emission from a star-forming galaxy is described in section
\ref{sec:method}. The model prediction is compared with the gamma-ray
observations of nearby galaxies in section \ref{sec:nearby}. Before
discussing the cosmic neutrino background, we examine the expected
contribution to IceCube neutrinos from the Galactic disk in section
\ref{sec:MW}.  Then the results on the cosmic gamma-ray and
neutrino background will be presented in section \ref{sec:galaxies}.
Discussions on model dependence and uncertainties are given in section
\ref{sec:uncertainty}, followed by summary in section
\ref{sec:summary}. We adopt a flat $\Lambda$CDM cosmology with
$\Omega_M=0.3$, $\Omega_B=0.05$ and $h=0.7$ throughout this work.

\section{Methods}
\label{sec:method}

\subsection{Production, Propagation and Interaction of Cosmic Rays}
\label{sec:CR}

Suppose that the four quantities of SFR (denoted as $\psi$), stellar
mass $M_*$, gas mass $M_{\rm gas}$, and the effective disk radius
$R_{\rm eff}$ (the radius including a half of the total galactic
light, related to the exponential scale $R_d$ as $R_{\rm eff} =
1.68R_d$) are given for a galaxy. We first need to determine the
production rate of pions by CR interactions as a function of the CR
proton energy $E_p$.

We assume that the production rate of CRs is proportional to SFR, and
the CR spectrum at the time of injection into the ISM is described by a
single power-law.  Then the number of CRs produced in a galaxy per
unit time and CR energy is expressed as
\begin{equation}
\label{eq:CR1}
\frac{dN_p}{dtdE_p} = C \left(\frac{\psi}{{\rm M}_\odot{\rm yr}^{-1}}
\right) \left(\frac{E_p}{{\rm GeV}}\right)^{-\Gamma_{\rm inj}},
\end{equation}
where the normalization factor $C$ will be fixed by fitting to the
observed gamma-ray luminosities of nearby galaxies later.
Observations of the cosmic-ray spectrum on Earth, gamma-ray spectrum
of supernova remnants and the Galactic diffuse gamma-ray emission
favor the injection spectral index in a range of $\Gamma_{\rm inj}=
2.2-2.4$ (see \cite{Ackermann12MW,Caprioli12} and references therein).
In this work we adopt $\Gamma_{\rm inj} = 2.3$ as the baseline value,
but we will also test the case of the strong shock limit, $\Gamma_{\rm
  inj} = 2$.

A fraction $f_\pi(E_p)$ of CRs interact with the ISM before they
escape into the intergalactic space, and it can be expressed as
$f_\pi(E_p)=1-\exp(-t_{\rm esc}/t_{\it pp})$, where $t_{\rm esc}(E_p)$
is the escape time of a CR particle from the galaxy and $t_{\it
  pp}(E_p)$ is the mean time to interact with the ISM by the
proton-proton ($pp$) collisions. This can be written as $t_{\it pp} =
(n_{\rm gas}\sigma_{pp} c)^{-1}$, where $n_{\rm gas}$ is the number
density of ISM and $\sigma_{\it pp}(E_p)$ is the inelastic part of the
$pp$ cross section, for which we use the formula given in
\citet{Kelner06}. We calculate $n_{\rm gas}$ as $n_{\rm gas} = M_{\rm
  gas}/(2 \pi R_{\rm eff}^2H_g m_p)$, where $m_p$ is the proton mass
and $H_g$ is the scale height of the gas disk.

Our galaxy formation model does not compute the scale height
$H_g$. The mechanism to determine disk heights of galaxies is complex
depending on many physical processes including magnetic field and
cosmic rays, gas pressure, and turbulent motion. In this work, we take
a simple empirical approach to estimate $H_g$ by assuming $H_g \propto
R_{\rm eff}$, and the coefficient is determined by the observations
for our Galaxy: $H_g = 150$ pc (\cite{Mo10}) and $R_{\rm eff} = $ 6.0
kpc (\cite{Sofue09}). Observations of nearby galaxies show that the
stellar scale height of disks ($H_*$) is roughly proportional to
$R_{\rm eff}$, and the above $H_g/R_{\rm eff}$ ratio is consistent
with observations if the difference between stellar and gas scale
heights is taken into account ($H_* \sim 2 H_g$ in our Galaxy).  In
section \ref{sec:uncertainty} we will also examine other modelings of
$H_g$ and effects on our results.  In the Mitaka galaxy formation
model, starbursts occur at the time of major merger of two disk
galaxies, resulting in a formation of spheroidal galaxy.  
We assume that such star-bursting galaxies are nearly spherical and
hence $H_g = R_{\rm eff}$.

The escape time $t_{\rm esc}$ is determined by the
shorter one of the CR diffusion time $t_{\rm diff}$ and the advection
time by galactic outflow $t_{\rm adv}$, i.e., $t_{\rm esc}=\min[t_{\rm
    diff},t_{\rm adv}]$.  These are estimated from galactic properties
as $t_{\rm diff}=H_g^2/[2D(E_p)]$ and $t_{\rm adv}=H_g/\sigma$, where
$D(E_p)$ is the diffusion coefficient of CRs and $\sigma$ is the
escape velocity from the gravitational potential of the galactic disk.
We estimate $\sigma$ from $H_g$ and 
the column density of total mass
$\Sigma = (M_* + M_{\rm gas})/(\pi R_{\rm eff}^2)$ 
assuming the relation for the 
isothermal sheet: $G \Sigma = \sigma^2/(2 \pi H_g)$ \citep{Mo10}.

The diffusion coefficient $D$ is not well constrained by
observations. Theoretically, $D$ is expected to depend on the CR
energy $E_p$ and magnetic field strength $B$.  Furthermore,
fluctuation pattern of magnetic fields is important for CR
diffusion. We consider two regimes regarding the proton Larmor radius $R_L
= 2.0 \times 10^{-7} (E_p/{\rm GeV})(B/{\rm 6 \ \mu G})^{-1}$ pc
following \citet{Aloisio04}. Suppose that there is a coherent length of turbulence $l_0$, and
magnetic fields are random and uncorrelated beyond $l_0$. This length
should be smaller than the region size, i.e., $l_0 \leq H_g$.  Then
the diffusion of high energy CRs with $l_0 < R_L < (H_g l_0)^{1/2}$ is
described by the small-angle random scattering approximation with a
mean free path $l_{\rm mfp} \sim (R_L^2/l_0)$, and hence $D \sim c
l_{\rm mfp}/3 \sim c R_L^2 / (3l_0)$.  It should be noted that when
$R_L > (H_g l_0)^{1/2}$ the mean free path becomes larger than $H_g$,
and hence in this regime we set $l_{\rm mfp} \sim H_g$, and hence $D
\sim cH_g/3$ and $t_{\rm diff} = 3H_g/(2c)$.  Diffusion of lower
energy CRs with $R_L < l_0$ is determined by resonant scattering in
the turbulent magnetic field fluctuations, resulting in $D \sim
cl_0(R_L/l_0)^\delta/3$, where $\delta$ depends on the spectrum of
interstellar turbulence and we adopt $\delta=1/3$ for the
Kolmogorov-type turbulence.  Then the diffusion coefficient can be
described as follows:
\begin{eqnarray}
\label{eq:diff}
D(E_p) = \left\{
\begin{array}{ll}
\frac{c l_0}{3} \left[ \left( \frac{R_L}{l_0}\right)^{\frac{1}{3}}
  + \left(\frac{R_L}{l_0} \right)^2 \right] 
& \left( R_L \leq \sqrt{H_g l_0} \right) \\
\frac{cH_g}{3}  &  \left( R_L > \sqrt{H_g l_0} \right)
\end{array}
\right.
\end{eqnarray}
(see also \cite{Parizot04}). 

The coherent length of turbulence $l_0$ is poorly understood
theoretically, but $l_{\rm outer} \sim$ 100--200 pc has been suggested
for turbulence generated by supernova remnants in the Milky Way
(\cite{Haverkorn08,Chepurnov10,Iacobelli13}), where $l_{\rm outer}$ is
the outer scale of turbulence and related as $l_0 = l_{\rm outer}/5$
for the Kolmogorov turbulence.  If this is a result of physics of
supernova remnants interacting with the ISM, it may not strongly
depend on the global properties of galaxies. Therefore, as a baseline
model we use a common value of $l_{0, \max}$ = 30 pc for all galaxies
whose $H_g$ is larger than $l_{0, \max}$ but $l_0 = H_g$ for smaller
galaxies, i.e., $l_0 = \min(l_{0, \max}, H_g)$.  Uncertainties and
other possible modelings about $l_0$ will be discussed section
\ref{sec:uncertainty}. The Larmor radius becomes larger
than $l_0$ for a very high energy of 
$E_p > 1.5 \times 10^{17} (l_0/{\rm 30 \ pc}) (B/{\rm 6 \ \mu G})$
eV, and hence the regime of $R_L < l_0$ is relevant for
GeV gamma-rays and PeV neutrinos in most galaxies. 

In order to estimate magnetic field strength, we assume that energy
density of magnetic field is close to that of supernova explosions
injected into ISM on the advection time scale, $t_{\rm adv} =
H_g/\sigma$. Then using a dimensionless parameter $\eta$, magnetic
field is given as $B^2/(8\pi) = \eta E_{\rm SN} r_{\rm SN} t_{\rm
  adv}/V$, where $E_{\rm SN}$ is the kinetic energy of a supernova,
$r_{\rm SN}$ the supernova rate in a galaxy, and $V = 2\pi R_{\rm
  eff}^2 H_g$ the volume of gas disk.  We set $\eta = 0.03$
because it reproduces the observed magnetic field strength of our
Galaxy (6 $\mu$G, \cite{Beck08,Haverkorn15}), using $\psi$, $R_{\rm
  eff}$, $M_{\rm gas}$ and $M_*$ in table
\ref{table:Fermi_galaxies}. Here we made standard assumptions that
stars heavier than $8 M_\odot$ end their life by suprenovae, $E_{\rm
  SN}=10^{51}\ {\rm erg}$ and the Salpeter IMF
(\cite{Woosley95,Salpeter55}) for all galaxies.

Now we can calculate the CR production rate $dN_p/dtdE_p$ and the
fraction $f_\pi$ of these interacting with the ISM as a function of $E_p$,
if $M_*$, $M_{\rm gas}$, $\psi$, and $R_{\rm eff}$ of a galaxy are
given.

\subsection{Neutrinos and Gamma-rays from Galaxies}
\label{sec:nu}

Then the number luminosity of neutrinos or gamma-rays,
i.e., number of particles produced per unit time and
particle energy in a galaxy is calculated as
\begin{equation}
\label{eq:gal}
\frac{dL_{N,i}}{dE_i} = \int_{E_i}^\infty f_\pi \frac{dN_p}{dtdE_p}
\frac{dn_i}{dE_i} dE_p,
\end{equation}
where the index $i$ denotes neutrino or gamma-ray ($i = \nu,\gamma$),
$dn_i/dE_i$ is the number of neutrinos or gamma-ray photons per unit
particle energy $E_i$ generated from one {\it pp} interaction, which
is a function of $E_i$ and $E_p$ and calculated by using the formulae
in \citet{Kelner06}. It should
be noted that in eq. (\ref{eq:gal}) we only consider the initial
$pp$ collision of each cosmic ray with the ISM, because a CR particle
loses a significant fraction of its energy in the first collision and
subsequent interactions have minor energetical contribution to the
gamma-ray or neutrino spectrum. Furthermore, the contribution
from the second and later interactions is effectively taken into
account because we fix the normalization factor $C$ to fit
the observed gamma-ray luminosity of nearby galaxies. 

The cosmic background radiation flux of gamma-rays or neutrinos 
per steradian and per unit $E_i$
is then calculated as
\begin{eqnarray}
\label{eq:nu_integ}
\Phi_i(E_i) &=& \frac{c}{4\pi} \int dz \left|
 \frac{dt}{dz} \right| (1+z) \frac{d{\cal L}_{N,i}
[(1+z)E_i; z]}{dE_i^{\rm rest}},
\label{eq:background}
\end{eqnarray}
where $E^{\rm rest}_i = (1+z) E_i$ is the rest-frame particle energy
and ${d{\cal L}_{N,i}[E_i^{\rm rest};z]}/{dE_i^{\rm rest}}$ is the
neutrino/gamma-ray number luminosity density at redshift $z$ per unit
comoving volume.  Our galaxy formation model generates mock galaxy
catalogues at various redshifts, in which the $j$-th galaxy is given a
weight $n_j^{\rm gal}$ (comoving number density of such a galaxy).  We
use morphology, $\psi$, $M_{\rm gas}$, $M_*$ and $R_{\rm eff}$ of
these catalog galaxies to calculate the number luminosity density 
by summing up the catalog galaxies as:
\begin{equation}
\frac{d{\cal L}_{N,i}[E_i;z]}{dE_i} = \sum_j n^{\rm gal}_j 
\left[\frac{dL_{N,i}(E_i)}{dE_i}\right]_j \ .
\end{equation}

For the gamma-ray background, additional corrections are necessary to
eq. (\ref{eq:background}) to take into account the attenuation of
gamma-ray flux due to $e^{\pm}$ creation by interaction with
background optical and infrared photons, and the subsequent cascade
emission produced by $e^\pm$s.  We use the baseline model calculation
of $\tau_{\gamma\gamma}[E_\gamma;z]$ in \citet{Inoue13} for the
absorption optical depth, and calculate the cascade emissivity
following the treatment of \citet{Inoue12}.

\subsection{Semi-Analytical Modelling of Galaxy Formation}
\label{sec:Mitaka}

In this work we use a semi-analytical model of cosmological galaxy
formation model presented by \citet{Mitaka}, which is called the
Mitaka model. This model first computes merger history of dark matter
haloes based on the extended Press-Schechter theory.  Then baryonic
processes in these haloes, such as gas cooling, star formation,
supernova feedback and galaxy mergers, are calculated with
phenomenological prescriptions. Free parameters included in describing
baryons are determined to match several observations of local
galaxies. This model reproduces various observed properties of local
and high-$z$ galaxies such as luminosity functions, size-luminosity
relation, luminosity densities, and stellar mass densities
(\cite{Mitaka,Kashikawa06,Kobayashi07},\yearcite{Kobayashi10}).  

This model includes two modes of star formation: starburst and
quiescent. Starburst is assumed to occur after major mergers of
galaxies leading to spheroidal galaxies, and otherwise stars are
formed in a disk (the quiescent mode).  Contributions of starbursts to
the total cosmic SFR in this model is about 5\% at $z\sim0$, which
increases to 15\% at $z\sim1$ and 30\% at $z\sim2$.

\section{Comparison with Nearby Galaxies}
\label{sec:nearby}

To fix the value of the normalization factor $C$ in the equation
(\ref{eq:CR1}), we use the data of six galaxies detected by the {\it
  Fermi}-LAT, including three normal galaxies (SMC, LMC, MW) and three
starbursts (NGC253, M82, NGC2146).  It should be noted that the latter
three galaxies are generally called ``starbursts'' by their intensive
star formation activity, but their morphology is disk-like without
evidence of major mergers in recent past. Therefore we treat these as
disk galaxies in theoretical modeling, and hence the definition of
"starburst'' is different from that (major merger) used in the galaxy
formation model.  Their physical quantities are shown in table
\ref{table:Fermi_galaxies}.  

There are several notes for this sample.  A different photon energy
range of gamma-ray luminosity $L_\gamma$ is used for NGC 2146 in the
literature, and hence we also change the range in theoretical
calculation accordingly.  Stellar mass $M_*$ for M82 is not yet
observationally well constrained, and we used a dynamical mass $M_{\rm
  dyn} \sim 10^{10} M_\odot$ \citep{Sofue92} and estimated $M_*$ from
$M_{*} = M_{\rm dyn} - M_{\rm gas}$. However, even if we set $M_* = 0$
for M82, our results are hardly changed. In addition to the galaxies
shown in table \ref{table:Fermi_galaxies}, M31, NGC4945, NGC1068 and
Arp220 are also detected in gamma-rays. However, we do not use
NGC4945, NGC1068 and Arp220 in our calculation because gamma-ray
emissions from these galaxies are likely affected by AGN activities
(\cite{Ackermann12SFG}, \cite{Yoast-Hull17}). M31 is also removed from
the sample, because a recent analysis of gamma-rays from M31 has shown
that the emission is not correlated with regions where most of the
atomic and molecular gas reside, suggesting that they are not
originated from the CR-ISM interactions (\cite{Ackermann17}).

\begin{table*}  
\caption{Properties of gamma-ray galaxies}
\begin{tabular}{ccccccccccc}
\hline \hline
Objects&$L_\gamma$&ref.&$\psi$&ref.&$M_{\rm gas}$&ref.&$M_*$&ref.&$R_{\rm eff}$&ref. \\ 
& [10$^{39}$ erg s$^{-1}$]$^*$ &&[$M_\odot$ yr$^{-1}$]&&[10$^9$ $M_\odot$]&&[10$^9$ $M_\odot$]&&[kpc]&\\ 
\hline	
MW		& 0.82$\pm$0.24	&(1)& 2.6			&(3),(4)& 4.9 	& (8)		& 50&		(14)& 6.0&	(19) \\ 
LMC		& 0.047$\pm$0.005	&(1)& 0.24 		&(4),(5)& 0.53   &(9)		& 1.5	&		(15)& 2.2&	(20) \\ 
SMC		& 0.011$\pm$0.003	&(1)& 0.037 		&(4),(5) & 0.45  &(10) 	& 0.46&		(15) & 0.7&	(21)\\
NGC253	& 6$\pm$2		&(1)& 7.9 			&(4),(6)& 4.3 	& (11)	& 21 &		(16)& 3.7&	(22)\\
M82		& 15$\pm$3 		&(1)& 16.3 		&(4),(6)& 1.3 	& (12)	& 8.7$^\dag$ & (17)& 1.2 &	(22)\\
NGC2146	& 40$\pm$21		&(2) & 17.5$^\ddag$ &(7)& 4.1 		&(13)	& 20 &		(18)& 1.8 &	(23)\\
\hline
\hline
\end{tabular}
\\ 
$^*$The photon energy range is 0.1--100 GeV except for
NGC 2146, for which 0.2-100 GeV is adopted according to
ref. (2). \\
$\dag$From a dynamical mass estimate (see text). \\
$\ddag$Estimated from 
the radio continuum luminosity at 1.4 GHz and eq. (13) of ref. (7). \\
References:
 (1) \citet{Ackermann12SFG}  (2) \citet{Tang14} (3) \citet{Diehl06} (4) \citet{Makiya11} (5) \citet{Kennicutt08} (6) \citet{Sanders03} (7) \citet{Yun01} (8) \citet{Paladini07} (9) \citet{Staveley-Smith03} and  \citet{Fukui08} (10) \citet{Stanimirovic99} and \citet{Leroy07} (11) \citet{Knudsen07} and \citet{Springob05} (12) \citet{Chynoweth08} and \citet{Mao00} (13) \citet{Tsai09} (14) \citet{Bland-Hawthorn16} (15) \citet{McConnachie12} (16) \citet{Lucero15} (17) \citet{Sofue92} (18) \citet{Skibba11} (19) \citet{Sofue09} (20) \citet{van_der_Marel06} (21) \citet{Gonidakis09} (22) J-band half-light radius in \citet{Jarrett03}, converted into kilopersec using distances taken from the NASA/IPAC Extragalactic Database (23) H$\alpha$ half-light radius in \citet{Marcum01}
\label{table:Fermi_galaxies}
\end{table*}

\begin{figure}
\begin{center}
\includegraphics[angle=270,width=\linewidth]{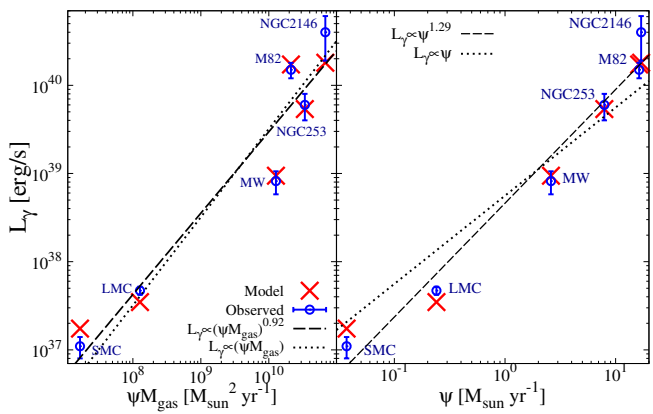}
\caption{Predicted (crosses) and observed (circles) gamma-ray
  luminosities of nearby galaxies are compared in $L_\gamma$ -- $\psi
  M_{\rm gas}$ (left panel) and $L_\gamma$ -- $\psi$ (right panel)
  plots.  The baseline model parameters described in the text are used
  in the model calculation.  The best-fit proportional (dotted) and
  power-law (dashed) relations to the observed data are also
  shown.  
}
\label{fig:fit}
\end{center}
\end{figure}

Then the parameter $C$ is determined by the standard $\chi^2$ fitting
to the six observed values of $L_\gamma$. Figure \ref{fig:fit} shows
comparison of gamma-ray luminosities between model predictions and
observations. We also show the best-fit proportional and power-law
relationships to the observed data as a function of $\psi$ or $\psi
M_{\rm gas}$, because fits to these quantities were often made in
previous studies, motivated by the expectation that CR production rate
is proportional to $\psi$ and the target gas mass for $pp$ reactions
scales with $M_{\rm gas}$.  Though our model is based on a simple
modeling about physical processes on the whole galactic scale, our
simple model nicely reproduces the observed luminosities, compared
with the power-law fits to $\psi$ or $\psi M_{\rm gas}$. It may be
rather surprising that this simple model predicts correct luminosities
for various types of galaxies, widely ranging from dwarf galaxies like
LMC and SMC to intense starburst galaxies.  Note that $C$ can be
converted to the energy injected into CRs from a supernova, as can be
seen from equation \ref{eq:CR1}. Setting the energy range of CRs to be
$10^9$--$10^{15}$ eV, the best-fit $C$ corresponds to CR energy of
0.24$E_{\rm SN}$ with the same $E_{\rm SN}$, IMF, and the threshold
stellar mass for supernovae assumed in the model.  This is comparable
to the standard assumption that CRs carry $\sim$ 10\% of supernova
explosion energy. This slightly larger value is partly a result 
of considering only the first time $pp$ interaction to produce
gamma-rays, as mentioned above.

\section{Gamma-rays and Neutrinos from the Galactic Disc}
\label{sec:MW}

\begin{figure}
\begin{center}
\includegraphics[angle=270,width=0.98\linewidth]{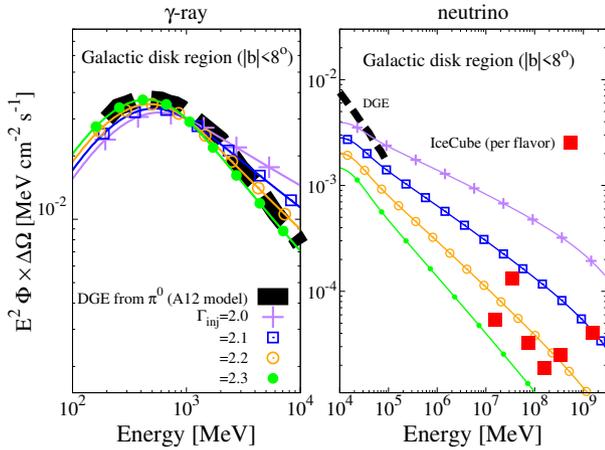}
\caption{Predicted gamma-ray (left panel) and neutrino (right panel)
  spectra from the Galactic disk region are shown. The spectral shapes
  of gamma-rays and neutrinos predicted by our model are shown with
  some different values of $\Gamma_{\rm inj}$.  The black dashed curve
  shows the spectrum of the diffuse Galactic gamma-ray emission (DGE)
  in the disk regions of $|b| < 8^\circ$, extracted from the {\it
    GALPROP}-based model of \citet{Ackermann12MW}.  The red squares
  are neutrino spectrum per flavor observed in all sky by the IceCube
  experiment, but multiplied by the solid angle fraction of the
  Galactic disk ($|$$b$$|$ $<$8$^\circ$) as an estimate for the
  observed neutrino flux in this region.}
\label{fig:MW}
\end{center}
\end{figure}

The gamma-ray sky measured by {\it Fermi} is dominated by the diffuse
Galactic emission (DGE), which is mostly from the pion decay, and
hence the Milky Way could also be a source of high-energy
neutrinos. It is important to check the predicted neutrino flux from
the Milky Way does not contradict with the IceCube data.  In figure
\ref{fig:MW} we show our model prediction of the Galactic diffuse
gamma-ray and neutrino emission with our baseline model parameters but
changing $\Gamma_{\rm inj}$.  For comparison, we also show the
pion-decay component of the $^{\rm S}$S$^{\rm Z}$4$^{\rm R}$20$^{\rm
  T}$150$^{\rm C}$5 model of DGE in \citet{Ackermann12MW} (A12 model,
hereafter) within the low latitude regions of $|b| < 8^\circ$, which
was calculated by the more detailed {\it GALPROP} code and is in good
agreement with the observed data.  The flux of our model is normalized
so that it matches the A12 model at GeV. We also show the IceCube
neutrino flux data, which is scaled into the $|b| < 8^\circ$ region
from the whole IceCube data assuming isotropy.

Figure \ref{fig:MW} indicates that our model matches the A12 model for
$\Gamma_{\rm inj}=2.3$, in which case the contribution from the Milky
Way can possibly explain 13\% of the IceCube energy flux. It should be
noted here that we do not introduce any cut off in the accelerated
cosmic ray spectrum (equation \ref{eq:CR1}), which may be rather
unrealistic.  Indeed, if we introduce a cutoff at PeV, for example,
the contribution decreases to 3\%.  Interestingly, some previous
studies have also suggested a considerable contribution to IceCube
neutrinos from the Milky Way
\citep{Gaggero15,Neronov15c,Palliadino16}. Increased statistics of the
IceCube neutrino events in the future may reveal an excess from the
Galactic disk region compared with the isotropic component, which
would give important information about the maximum cut-off energy of
proton acceleration in the Milky Way. On the other hand, a hard
injection spectrum of $\Gamma_{\rm inj} \lesssim 2.1$ is disfavored
because the Galactic disk component of neutrinos would be too strong,
unless there is a cut-off below $E_p \sim 10^{14}$ eV.

\section{Cosmic Gamma-Ray and Neutrino Background}
\label{sec:galaxies}

\begin{figure}
\begin{center}
\includegraphics[angle=270,width=\linewidth]{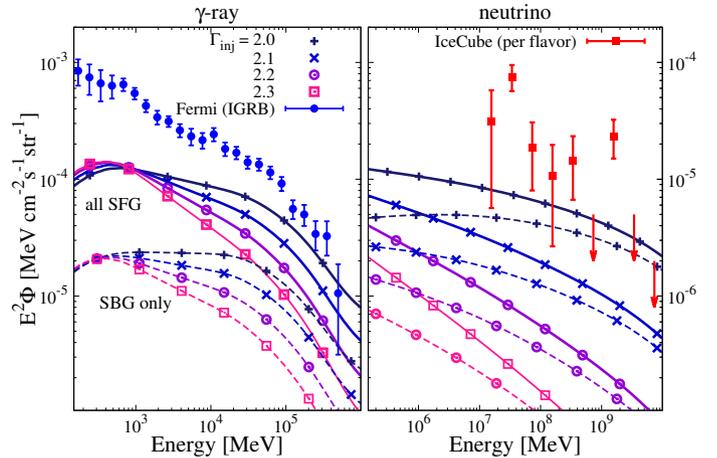}
\caption{The cosmic diffuse background of gamma-rays (left panel) and
  neutrinos (right panel) from star-forming galaxies predicted by our
  baseline model are shown for different values of $\Gamma_{\rm
    inj}$. Solid and dashed lines correspond to the contributions from
  all galaxies and starburst galaxies respectively.  Data points
  represent the gamma-ray spectrum of unresolved isotropic gamma-ray
  background observed by {\it Fermi}-LAT (blue, \citet{IGRB}) and the
  astrophysical neutrino spectrum per flavor observed in the IceCube
  experiment (red, \citet{IceCube}). For the purpose of presentation,
  the scale of vertical axis is different between the left and right
  panels.}
\label{fig:baseline}
\end{center}
\end{figure}

Figure \ref{fig:baseline} presents the cosmic gamma-ray and neutrino
background spectra predicted by our baseline model, in comparison with
the {\it Fermi} and IceCube data.  Our calculations show that the
gamma-ray energy flux from star-forming galaxies is ($5.1$ --
$7.0)\times 10^{-4}$ MeV cm$^{-2}$s$^{-1}$str$^{-1}$ above 100 MeV,
which is 18 -- 25\% of the IGRB flux observed by {\it Fermi}, in
reasonable agreement with previous studies
(\cite{Fields10,Makiya11,Stecker11,Ackermann12SFG,Lacki14}).

The neutrino flux predicted by our baseline model ($\Gamma_{\rm inj}$
= 2.3) is only 0.5\% contribution to the IceCube data. If we assume a
harder spectrum at injection $\Gamma_{\rm inj} = 2.1$ and 2.2, the
contributions increase to 8.4\% and 2.1\% respectively.  Even in the
most optimistic (and extreme) case of $\Gamma_{\rm inj} = 2$ in all
galaxies, star-forming galaxies can account for only 22\% of the
IceCube data.  It should be noted that such a hard injection spectrum
in our Galaxy is in contradiction with the Galactic diffuse gamma-ray
spectrum and the isotropy of the IceCube data.  Therefore, we conclude
that star-forming galaxies cannot be the major source of the IceCube
neutrinos, and a reasonable estimate of their contribution is about
1-8\% or less.

\section{Dependence on model parameters}
\label{sec:uncertainty}

\begin{figure}
\begin{center}
\includegraphics[angle=270,width=0.95\linewidth]{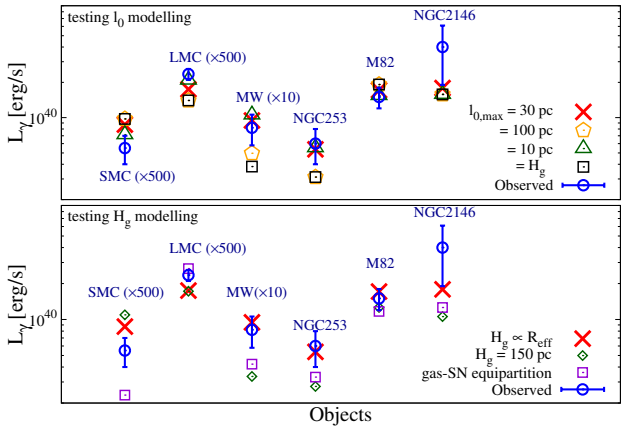}
\caption{Comparisons between predicted and observed gamma-ray
  luminosities of nearby galaxies for different modellings of $l_{0,\max}$
  (upper panel) and $H_g$ (lower panel).  In both panels the
  red cross points correspond to our baseline model. For the purpose
  of presentation, gamma-ray luminosities of SMC, LMC, and MW are
  multiplied by some numbers indicated in the figure.
  }
\label{fig:fit2}
\end{center}
\end{figure}

\begin{figure}
\begin{center}
\includegraphics[angle=270,width=0.95\linewidth]{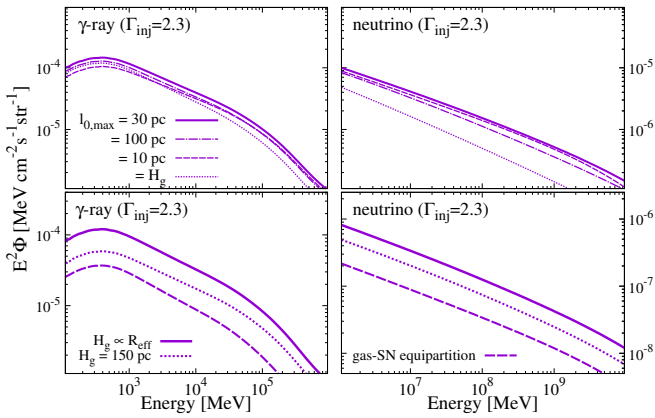}
\caption{Same as figure \ref{fig:baseline}, but showing dependence on
  the modelling of $l_{0,\max}$ (upper panels) and $H_g$ (lower
  panels), using the baseline model value of $\Gamma_{\rm inj} = 2.3$.
}
\label{fig:model}
\end{center}
\end{figure}

Dependence of our results on different modelings of $l_{0,\max}$ and
$H_g$ is shown as the change of the gamma-ray luminosities of nearby
galaxies (figure \ref{fig:fit2}) and the cosmic gamma-ray/neutrino
background flux (figure \ref{fig:model}) from our baseline model.  For
$l_{0,\max}$ we test different values of 10 and 100 pc from the
baseline model (30 pc), and also test a model with $l_{0,\max} = H_g$,
as the case where the coherent scale of turbulence is determined by
the system size.  The model with $l_{0,\max}$ = 10 pc also agrees
reasonably well with the $L_\gamma$ data of nearby galaxies, and the
models with $l_{0,\max}$ = 100 pc and $l_{0,\max}=H_g$ show larger
deviation, especially MW and NGC 253 up to a factor of 3. The changes
of background fluxes are small (less than 20\%) for $l_{0,\max} = $ 10
or 100 pc, but the neutrino flux is reduced by a factor of 2--3 in the
model of $l_{0,\max}=H_g$. This is because $l_0$ becomes larger
in galaxies of $H_g > $ 30 pc, and the diffusion coefficient
becomes larger with larger $l_0$ if $R_L < l_0$ which is
valid in most galaxies.

To check the dependence on the modelling of $H_g$, we test the
following two cases for disk galaxies: (1) assuming that the height of
gas disk in all disk galaxies are similar to that of the Milky Way,
i.e., $H_g$ = 150 pc, and (2) assuming energy equipartition between
gas motion and energy produced by supernovae, as $\rho_{\rm
  gas}\sigma^2 = \alpha E_{\rm SN}r_{\rm SN}t_{\rm adv}/V$, where
$\rho_{\rm gas} = M_{\rm gas}/V$ is ISM gas density. In the latter
case both $\sigma$ and $H_g$ are determined by this relation combined
with $G \Sigma = \sigma^2/(2 \pi H_g)$, without using $H_g \propto R_{\rm
  eff}$.  The dimensionless factor $\alpha = 4$ is determined to
reproduce $H_g$ = 150 pc for physical quantities of the Milky Way.

The lower panel of figure \ref{fig:fit2} shows that different
modelling of $H_g$ results in clear discrepancies between predicted
and observed $L_\gamma$ for local galaxies. As shown in the lower
panels of figure \ref{fig:model}, the gamma-ray and neutrino
background flux is reduced in the constant $H_g$ model by a factor of
about 2. This is likely because the gas density is underestimated in
small galaxies whose density is higher if we assume $H_g \propto
R_{\rm eff}$. The background fluxes are reduced even by a larger
factor of about 3--4, for the model assuming equipartition between gas
and supernovae. This implies that gas density is higher than that
expected from equipartition. Such a situation may be expected,
especially in starburst galaxies, if gas density becomes high enough
before star formation starts and supernova energy is injected to the
ISM. Figure \ref{fig:fit2} indeed indicates that the discrepancy in the
gas-SN equipartition model is larger for starburst galaxies.  In any
case, different modelings of $l_{0,\max}$ and $H_g$ lead to
lower neutrino background flux, and hence it does not affect our
conclusions that the majority of IceCube neutrinos cannot be explained
by star-forming galaxies.

\section{Summary}
\label{sec:summary}

In this work we constructed a new theoretical model to predict flux
and spectrum of gamma-rays and neutrinos by interactions of
cosmic-rays produced by supernovae in a star-forming galaxy, and
applied it to predict the flux and spectrum of the cosmic gamma-ray
and neutrino background radiation.

Our model calculates gamma-ray and neutrino spectra of a star-forming
galaxy from four physical quantities, i.e. SFR, size, gas
mass, and stellar mass, taking into account the production, propagation
and interactions of cosmic rays in the ISM. This model is tested against
the gamma-ray luminosities measured by {\it Fermi} of the six nearby
galaxies listed in table \ref{table:Fermi_galaxies}.  In spite that
this sample includes a wide variety of star-forming galaxies, from
dwarf galaxies like LMC and SMC to starburst galaxies such as M82, our
model reproduces the observed gamma-ray luminosities fairly well,
within a factor of $\sim 2$. The agreement is
significantly better than the simple phenomenological modelings
assuming a power-law relation between $L_\gamma$ and SFR or SFR$\times
M_{\rm gas}$. This model can be tested with a larger sample
of nearby galaxies by high-sensitivity gamma-ray observations 
of future projects such as Cherenkov Telescope Array (CTA). 

We have examined the neutrino emission from the disk of our Galaxy
predicted by our model, and found that the observed isotropic
distribution of IceCube neutrinos constrains the injection cosmic-ray
energy spectrum to be softer than $\Gamma_{\rm inj} \sim 2.2$,
unless there is a cut-off at $\lesssim 10^{14}$ eV in the CR energy
spectrum. 

The contribution from star-forming galaxies to the extragalactic
gamma-ray and neutrino background is calculated by using a
semi-analytical cosmological galaxy formation model. This 
 model is quantitatively in agreement with many observations
of galaxies at local and high redshifts.  We have found that
star-forming galaxies make about 20 \% of the isotropic gamma-ray
background flux unresolved by {\it Fermi}, and only 1-8\% or less of
the IceCube flux, with reasonable values of $\Gamma_{\rm inj} =
2.1-2.3$. Even with the most optimistic case where $\Gamma_{\rm inj} =
2.0$ in all galaxies, the contributions is no more than
22\%. Therefore, we conclude that the majority of IceCube neutrinos
cannot be explained by star-forming galaxies.  We also examined
dependence of these results on modelings about $l_{0, \max}$ (the
maximum length of coherence in ISM turbulence) and $H_g$ (gas scale
height of a galactic disk), and found that alternate prescriptions
give even lower neutrino flux.

Our results demonstrate that star-forming galaxies make only a minor
contribution to the IceCube flux, implying that other sources are
required to explain most of the observed data. Since correlation
analyses have shown that gamma-ray blazars and gamma-ray bursts are
not the dominant source (\cite{IceCube15}, \yearcite{IceCube17}),
there is no strong candidate for the origin of IceCube neutrinos.
Future detectors such as the IceCube-Gen2, as well as multi-messenger
approaches with next-generation telescopes like CTA, will be the key
to understand the nature of high-energy neutrinos.

\begin{ack}
TT was supported by JSPS KAKENHI Grant Numbers JP15K05018 and JP17H06362. NK is supported by the Hakubi project at Kyoto University.
\end{ack}

\end{document}